# Performance Results of HESP Physical Model

Anantha Chanumolu[2*] . Sivarani Thirupathi[1] . Damien Jones[3] . Sunetra Giridhar[1] . Deon Grobler[4] . Robert Jakobsson[4] .

[1] *Indian Institute of Astrophysics, Bangalore India 560034*
[2] *Max Planck Institute for Solar System research, Göttingen, Germany 37075*
[3] *Prime Optics, Eumundi, Australia Q4562*
[4] *Kiwistar Optics, Lower Hutt, New Zealand 5040*

*Email:* chanumolu@mps.mpg.de
sivarani@iiap.res.in

**Abstract** As a continuation to the published work on model based calibration technique with HESP(Hanle Echelle Spectrograph) as a case study, in this paper we present the performance results of the technique. We also describe how the open parameters were chosen in the model for optimization, the glass data accuracy and handling the discrepancies. It is observed through simulations that the discrepancies in glass data can be identified but not quantifiable. So having an accurate glass data is important which is possible to obtain from the glass manufacturers. The model's performance in various aspects is presented using the ThAr calibration frames from HESP during its pre-shipment tests. Accuracy of model predictions and its wave length calibration comparison with conventional empirical fitting, the behaviour of open parameters in optimization, model's ability to track instrumental drifts in the spectrum and the double fibres performance were discussed. It is observed that the optimized model is able to predict to a high accuracy the drifts in the spectrum from environmental fluctuations. It is also observed that the pattern in the spectral drifts across the 2D spectrum which vary from image to image is predictable with the optimized model. We will also discuss the possible science cases where the model can contribute.

---

\* *The work was carried out by the author during her time at the Indian Institute of Astrophysics*

# 1 Introduction

A modelling scheme that predicts the centroids of spectral features of a high resolution Echelle spectrograph[1] to a high accuracy and the preliminary results of its performance were presented in [1]. As discussed in [1], once the model is built, the next step is to adjust various parameters in the model to match it with the built instrument. Shown in Fig 1.1, is the flow chart of the procedure to match the model and the built instrument. The model maps predefined slit position on to a final position on the CCD, for a set of accurate Thorium Argon wavelengths and a set of parameters. The parameters are physical distances and angles in the instrument. The ThAr calibration exposures that are used for the wavelength calibration of the science data will be used to match the model and the instrument. The centroid positions of a set of ThAr features will be measured and matched with the predicted positions of the model by adjusting the open parameters in the model. A merit function (generally root mean square error) is constructed to minimize the difference between the model predictions and the measured centroids of the ThAr features from the calibration frame. To adjust the parameters a global optimizer, simulated annealing is used.

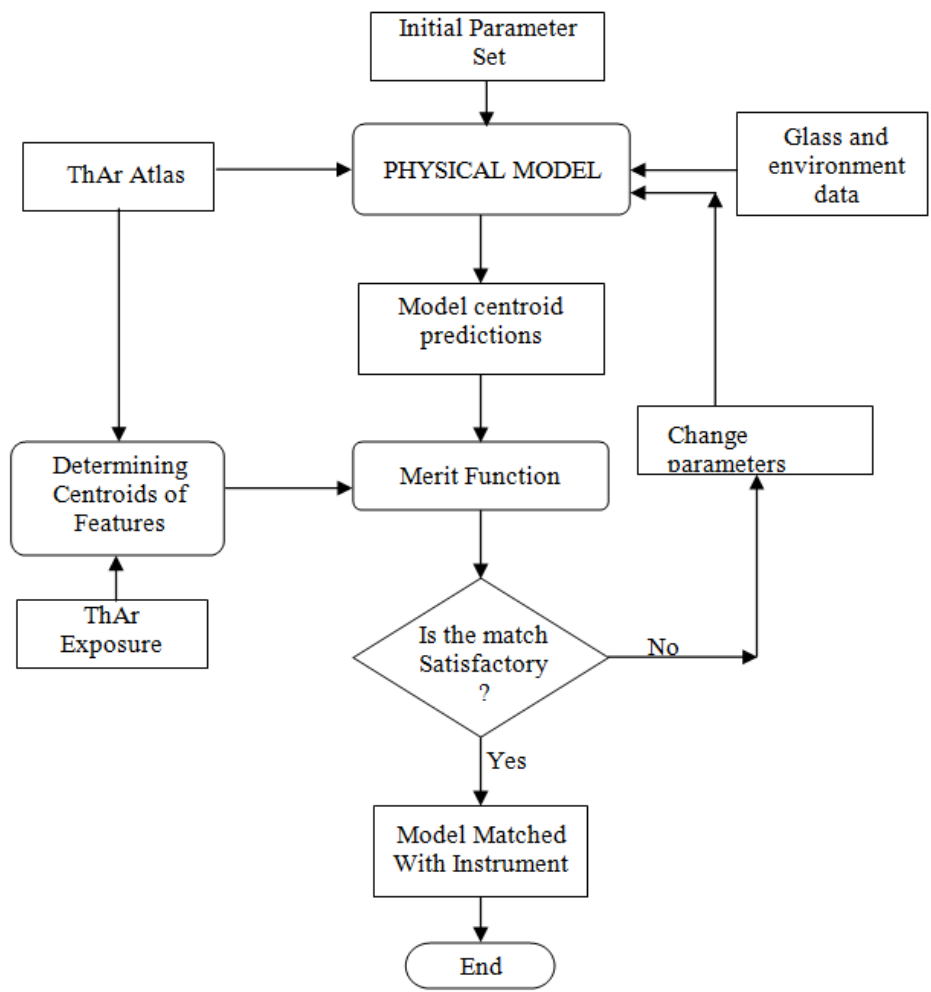

**Fig 1.1** Flow chart of model optimization to match with the built instrument [2]

Apart from accurate modelling of the instrument, the supplementary data (parameter set to optimize, glass and environmental data, accurate measurement of features' centroids in the calibration frame, accuracy of wave lengths from ThAr atlas) to be fed to the optimiser is also very crucial. We had an advantage of developing the model for the Hanle Echelle Spectrograph (HESP), when the subsystems of the instrument were being built. This gave a good understanding of the built optics and opto-mechanics of the instrument. The measured dimensions of various components are used in the model.

Sections 2 and 3 discuss the choice of parameter set and the effect of discrepancy in glass data on the optimization. The choice of wavelength calibration features, their centroid determination and the optimization procedure were discussed in [1]. The model based techniques performance tests are presented in section 4.

## 2 Choice of open parameters for optimization

Various physical parameters - such as distances between the components, tilts associated with various surfaces, decenters in the positions- form the parameter set for the model. The initial instrument model used design parameters. A total of 45 parameters were identified in HESP model: Slit's x, y and z-tilts, slit x & y decentres, slit to collimator distance, collimator's x and y tilts, Collimator to Echelle distance, Echelle x, y and z- tilts, Echelle grating constant, Collimator to slit mirror distance, Fold mirror x and y-tilt, x and y decentres for prism1, fold mirror to prism distance, prism 1 y-tilt, prism 2 decentre, prism 1 to prism 2 distance, camera x and y decentre, prism 2 to camera distance, y-tilts before and after prism2, prism1 and 2 x-tilt, camera to filed flattener distance, Filed flattener and CCD x and y decentre, x, and z tilts of field flattener and CCD , field flattener to CCD distance, CCD x and y decentres, CCD x, y and z tilts.

These values may vary slightly from the instrument design numbers once the instrument is built and aligned. So it is necessary to establish the exact values. This may also be the case after a major maintenance intervention, upgrade to the instrument or even an earthquake, resulting in a physical change in the instrument. A subset of these parameters will be optimized using the calibration images. Out of the total 45 parameters, 25 are used in optimization. The effect of change in various parameters at the detector plane for different wavelengths is studied using patterns in the vector maps of differences between the design and perturbed model's spectral positions. Fig 2.1 shows some examples of vector maps showing the differences in positions of selected spectral line features when various parameters are perturbed within acceptable limits. Dots represents the spectral features across the detector location as per the design (nominal positions) and the arrow of vectors indicate drift in the features with change in parameters.

The final shifts result from the combined effect of multiple parameters. In order to understand the parameters that need to be optimized, various subsets of parameters are selected for optimization. Studying the residuals of these various cases, the final subset of parameters for optimization can be chosen. For example, though the major shifts in the high dispersion and cross dispersion directions (high dispersion is the dispersion axis of the grating while the cross dispersion is the prism dispersion axis) can be attributed to Echelle and fold mirror angles, the second and higher order shifts cannot be attributed to the same. An example is shown in Fig 2.2.

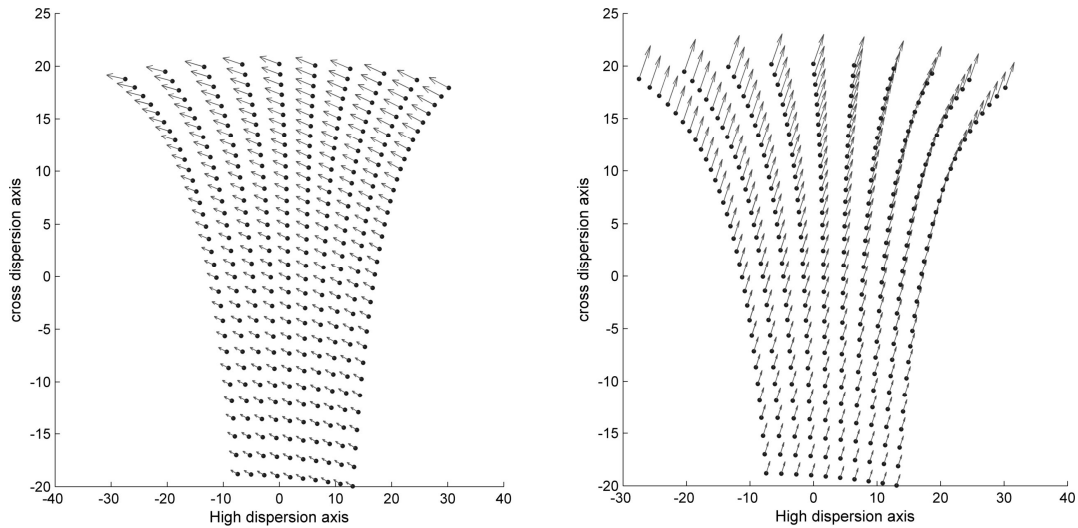

**Fig 2.1** Shift in spectrum for different parameter values. Two different parameter subsets were changed a little from the nominal. The dots show the scatter of selected wavelengths across the spectrum and the arrows indicate the shift in these positions from the nominal when the parameter subsets are changed for two different parameter subsets (The length of arrow is scaled according to the minimum and maximum of the shifts and is not equal to the axes scale).

A ThAr spectrum is obtained from the aligned instrument and only few basic parameters like the Echelle x-tilt, fold mirror y-tilt, slit decentre, collimator and prism tilts were optimized. Fig.2.2 shows the vector plots for residuals before and after optimization with different selected parameters. There is a possibility of some parameters compensating for others causing degenerate states. While these will be hard to resolve, they can be minimized by understanding the instrument alignment and its components' manufacturing procedures. The measurements of the manufactured and aligned optical and mechanical components can correct for maximum differences, and an understanding of the operation of the instrument and the conditions in which it operates can dictate the bounds on the parameters for optimizing and hence minimising the number of degeneracy states.

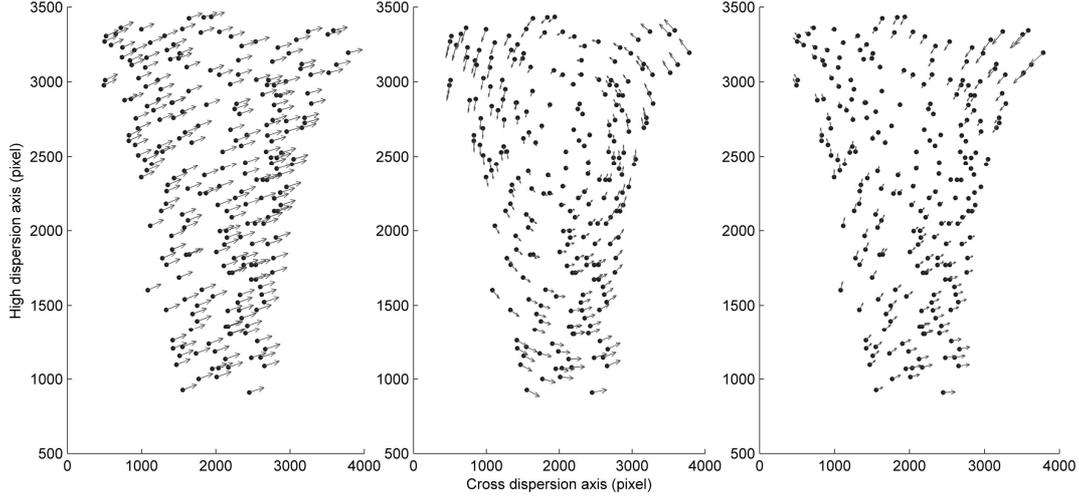

**Fig 2.2** Left: Vector plot of difference of selected features' positions from instrument and model predictions before optimization. Centre & Right: Vector plots of difference of features' positions from instrument and optimized model predictions (model optimized for two different subsets of parameters). The length of vectors is scaled according to the minimum and maximum of the shifts and is not equal to the axes scale. The residual patterns are an indication of missing parameters used for optimization. The centre plot rotational pattern of the residuals indicate the a major discrepancy in the field flattener tilts, and the right plot's pattern indicate a discrepancy in Echelle y-tilt

# 3 Glass refractive indices

In HESP, prism and the camera optics are made of various glasses from OHARA and SCHOTT. While BSL7Y was used for the cross disperser prisms, S-FPL51Y, BAL15Y, S-FSL5Y, BSL7Y, N-BAK2, Fused SILICA were used for camera optics. Optical properties of relevance for a glass are – refractive index, dispersion formula, and effect of temperature on refractive index. Any glass manufacturer's catalogues provide all the required information. Since, we did not have the facilities to carry out these measurements; we had to rely on the methods and results from the manufacturer. Refractive index of glass is measured using either v-block refractometer which gives an accuracy of about +3E-5 or with precision spectrometers that can measure indices at an accuracy level of 4E-6. Usually, manufacturers provide refractive index values for some wavelengths. In order to calculate the refractive index at any wavelength, a dispersion formula is defined. There are many formulas that fit the dispersion but the most generally used is the Sellmeier formula,

$$n^2 - 1 = \frac{A_1 \lambda^2}{\lambda^2 - B_1} + \frac{A_2 \lambda^2}{\lambda^2 - B_2} + \frac{A_3 \lambda^2}{\lambda^2 - B_3} \qquad (1)$$

where $A_1, A_2, A_3, B_1, B_2,$ and $B_3$ are the constants that can be found by least square fitting using the refractive indices of standard wavelengths measured from several melt samples and $\lambda$ is wavelength in micrometers. Using formula (1) the refractive index of any wavelength can be found to an accuracy of about $\pm$5E-6. Refractive index is affected by change in temperature. Schott Glass Technologies Inc. has developed a model for calculating the change in refractive index with temperature described in [3]. Manufacturers can provide various required measurements on individual melts on request. In order to match model with the instrument, it is important to get the refractive index data correct. Since the glass blanks were purchased much before this work started, we did not get a chance to request the glass manufacturers for the coefficients and data of the melt. The refractive indices and dispersion will vary slightly from melt to melt, as these are highly dependent on annealing factors. Using the standard melt data from the glass manufacturer (refractive indices for selected wavelengths) the coefficients of the dispersion formula are re-optimized. If the complete data of the melt was obtained from the manufacturer a more accurate dispersion formula is possible. In order to understand the implications of difference in coefficients from actual values, some simulations were performed. Prism is responsible for the cross dispersion of orders. A difference in the dispersion formula will cause a difference in the order spacing and to very small extent order tilts. Since we are using only four to six data points to adjust the dispersion solution, two different optimizations were tried,

1) $A_1$, $A_2$, $A_3$ were optimized to fit to the data keeping $B_1$, $B_2$, $B_3$ constant

2) $B_1$, $B_2$, $B_3$ were optimized to fit to the data keeping $A_1$, $A_2$, $A_3$ constant

to induce a difference in dispersion formula. Fig 3.1 shows the difference in the refractive indices for the two dispersion formulae which are the order of $10^{-5}$.

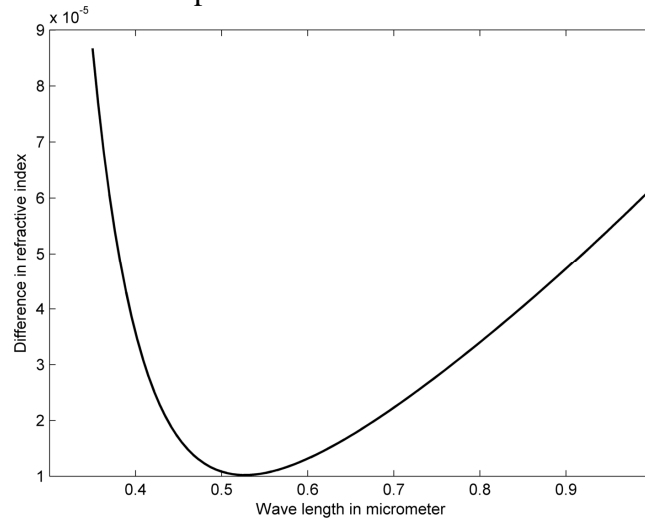

**Fig 3.1** Difference in the refractive index for two optimizations described in text

Fig 3.2 shows the shift in the spectrum between the two dispersion formulae for the prism glass in HESP spectrograph, assuming the values used for the glasses in the camera optics is accurate. It will be difficult to isolate this effect from the rest of the parameters in the model. Also during optimization, this effect may get compensated to an extent by other parameters. Hence, the best choice will be to get the most precise melt data from the glass manufacturers or make measurements in a proper facility.

Fig 3.3 shows pattern in the shift in spectrum with a little alteration of the dispersion coefficients in camera optics. With the discrepancies from measurements and fitting errors kept within possible limits, the spectrum shifts were well within one pixel. In the Fig 3.3 it can be seen that for various differences introduced, the pattern introduced is similar with varying magnitudes.

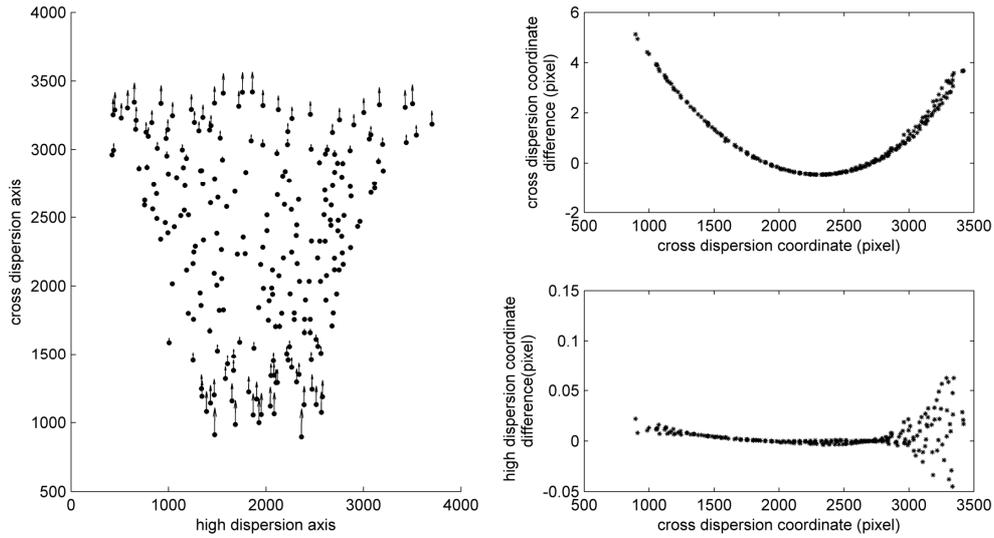

**Fig 3.2** Depiction of shift in spectrum from calculated positions due to mismatch in prism glass refractive indices data. Left: Vector plot showing the pattern in shift, Right: Shift in cross and high dispersion coordinates with respect to cross dispersion coordinate of the wavelengths positions on the CCD

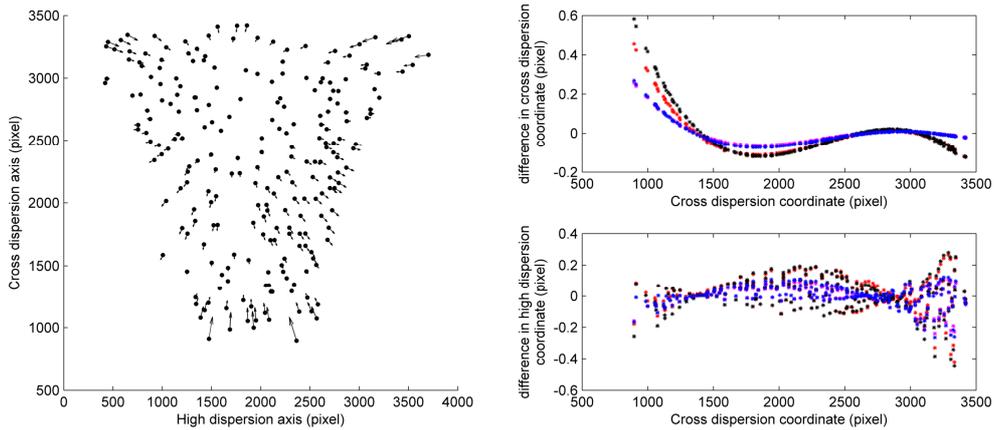

**Fig 3.3** Depiction of shift in spectrum from calculated positions due to mismatch in camera glasses refractive indices data. Left: Vector plot showing the pattern in shift, Right: Shift in cross and high dispersion coordinates with respect to cross dispersion coordinate of the wavelengths positions on the CCD for multiple cases

Spectrum taken from the instrument is a result of all the discrepancies in the system which are not easy to separate. These differences will be treated as a polynomial correction at the detector plane. This will be discussed in the results section.

In order to understand how the optimization will perform due to mismatches in glass data, a simulation was done. The constants in the dispersion formulae of the glass data were changed randomly by a small amount and the positions of the features in the spectrum were calculated. In Fig 3.4 the residuals for two cases of optimizations was shown. + sign shows the residuals after optimization when the glass data of the model and instrument are same. O sign shows the residuals when the glass data in model and actual instrument are different. The residuals from the later case are not totally compensated by the parameters used for optimization. It is observed that the discrepancy due to glass data of the prism is compensated to some extent by other system parameters, while large residuals are still present showing the effects from the inaccurate camera optics' glass dispersion (which is concluded by simulating data sets for refractive index discrepancy in prism and camera glasses separately and optimizing the model parameters for so obtained data sets). Hence a discrepancy can be detected due to the inaccurate input glass data, however the actual contribution from each component cannot be worked out.

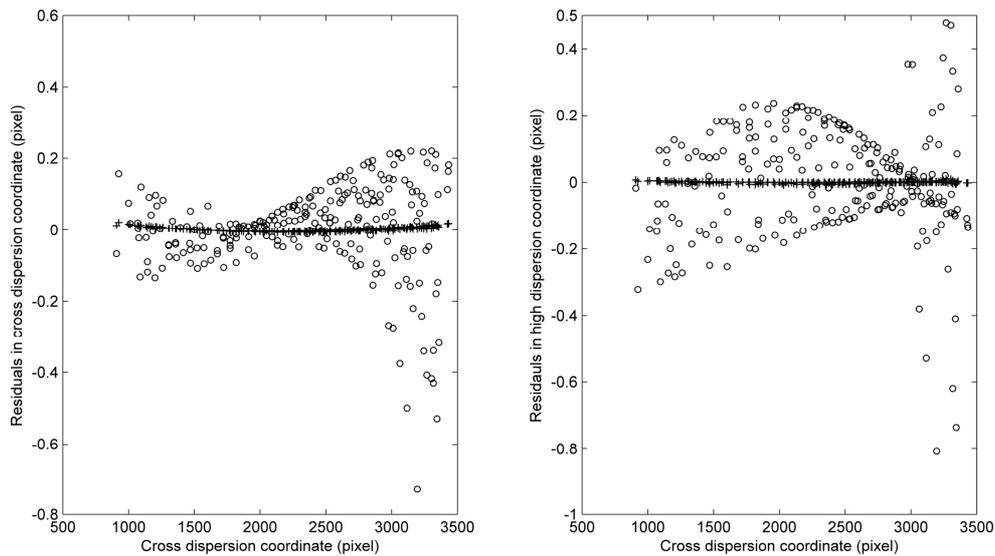

**Fig 3.4** Residuals in the wave lengths positions on the CCD after optimization of the model for two cases. +: when the glass data of the model and instrument matched, o: glass data of model and instrument differ

## 4 Results

Weighted root mean square error is chosen as the merit function for optimization, which is the euclidean distance between the measured positions of the wavelength features from the instrument and the model predictions. Different Th lines are of different intensities. The reciprocal of uncertainty in measured centroids of the Th features are the weights so that the features with less error are given more preference in the merit function. A set of 550 wavelength features covering all the orders of the spectrum (not necessarily uniformly distributed), with good intensity levels are chosen and model is optimized. Fig 4.1, 4.2 shows the scatter of residuals of this set (difference in measured centroids from the calibration frame and the optimized model predictions).

In Fig 4.1, the high dispersion coordinates residuals versus cross dispersion coordinate plot, the residuals are large at the far left end, which is in agreement with the low intensity in the blue end of the spectrum. The number of lines selected in different orders is different. While some mid orders have large numbers of usable lines, blue and red ends have few usable features. To understand if the number of features in different orders has any effect on optimization of the model, almost same number of features were selected (233 in total) in each order (four or less features per order) and model is optimized. Fig 4.3 shows the residuals from this optimization.

From the scatters it is obvious that the residuals are not affected by choosing less number of lines per order as long as they cover the spectrum. In principle it is not necessary to select features in every order, but to optimize the model to a good accuracy including the nonlinearities, it is good to choose lines covering the edges and the centre regions considering there are no discontinuities in the dispersion and distortions in the spectrum. 720 features which are not used for optimization were used to test the accuracy of the optimized model. Scatter of residuals (blue points) in the Fig 4.4 shows the performance of the optimized model. A low order 2D polynomial is fit to the residuals to take care of the patterns observed in Fig 4.3. Green scatter in Fig 4.4 shows the residuals after the polynomial fit. The box around the scatter depicts one pixel size.

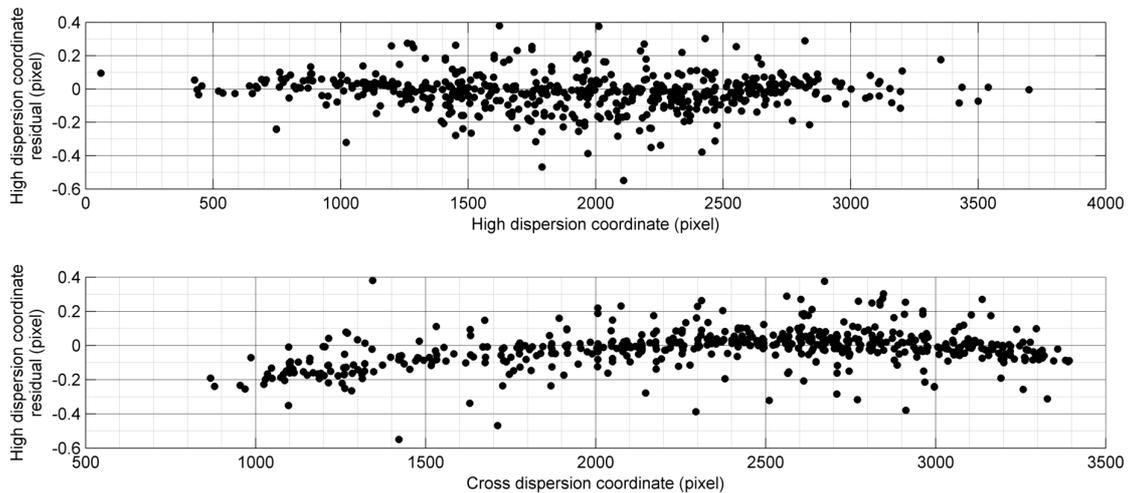

**Fig 4.1** Residuals of the high dispersion axis coordinates of 550 features used for optimization with weighted mean square as the merit function. Top panel: Difference between measured and optimized model predicted high dispersion axis coordinates versus High dispersion coordinate of the respective feature, Bottom Panel: Difference between measured and optimized model predicted high dispersion axis coordinates versus Cross dispersion coordinate of the respective feature

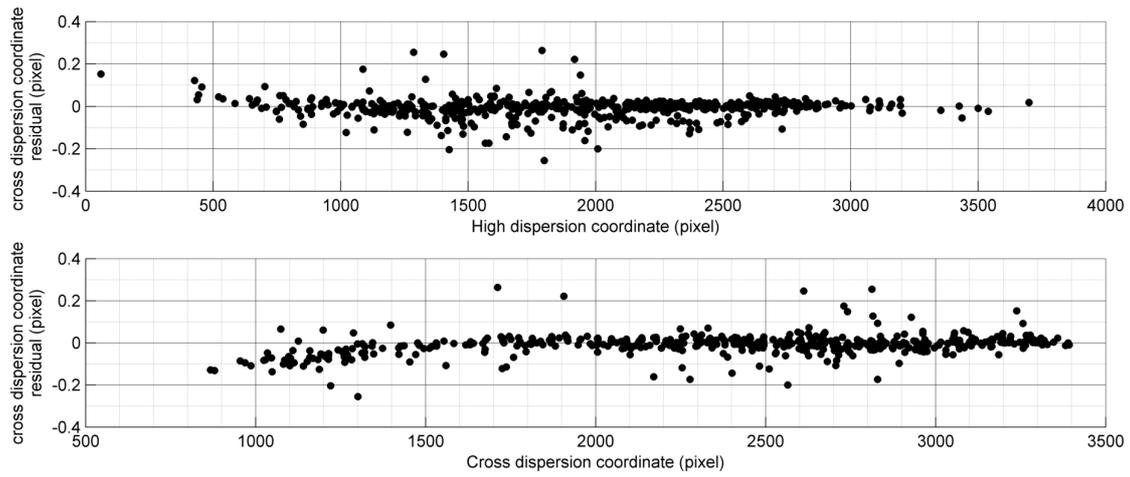

**Fig 4.2** Residuals of the cross dispersion axis coordinates of 550 features used for optimization with weighted mean square as the merit function. Top panel: Difference between measured and optimized model predicted cross dispersion axis coordinates versus High dispersion coordinate of the respective feature, Bottom Panel: Difference between measured and optimized model predicted cross dispersion axis coordinates versus Cross dispersion coordinate of the respective feature

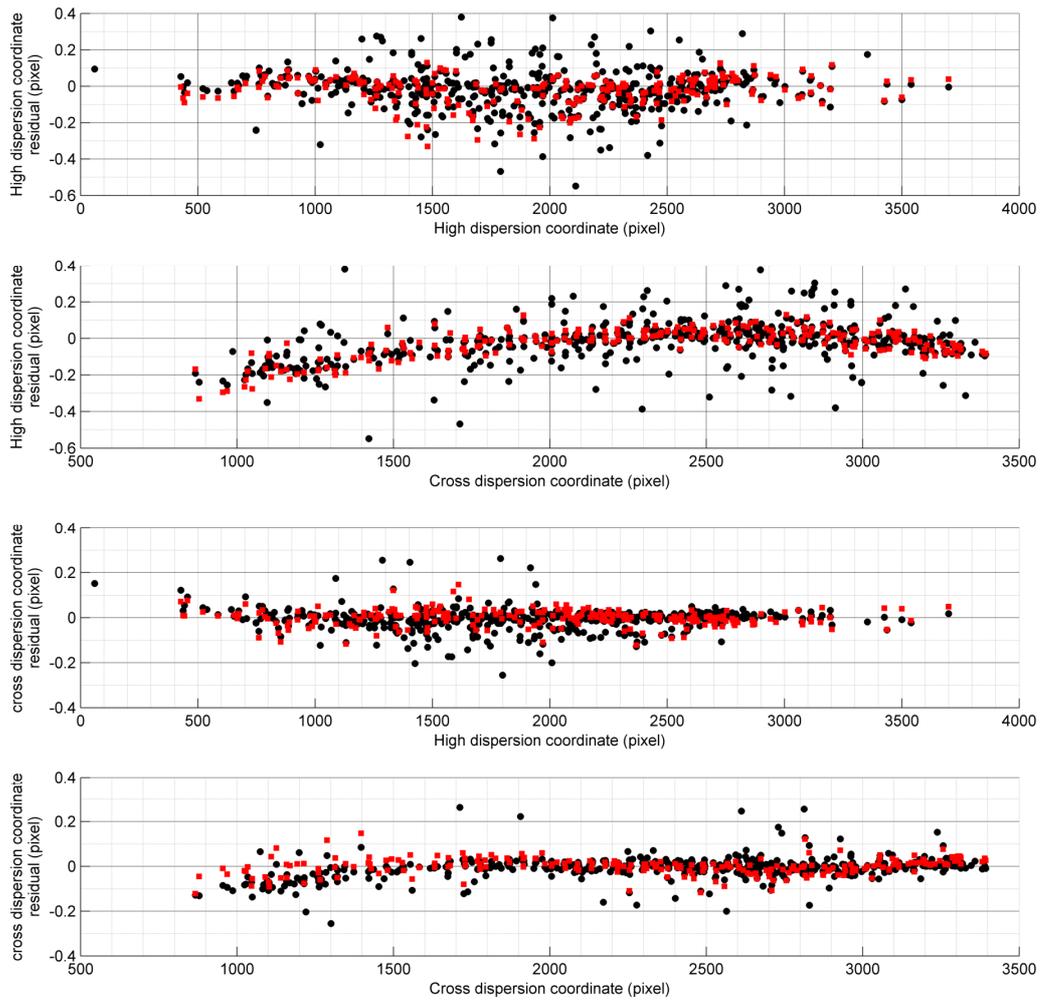

**Fig 4.3** Black scatter points are same as Fig 4.1 and Fig 4.2. These scatters are over plotted with the respective residuals of the 233 features (a subset of the 550 features) after the model was optimized using the 233 features

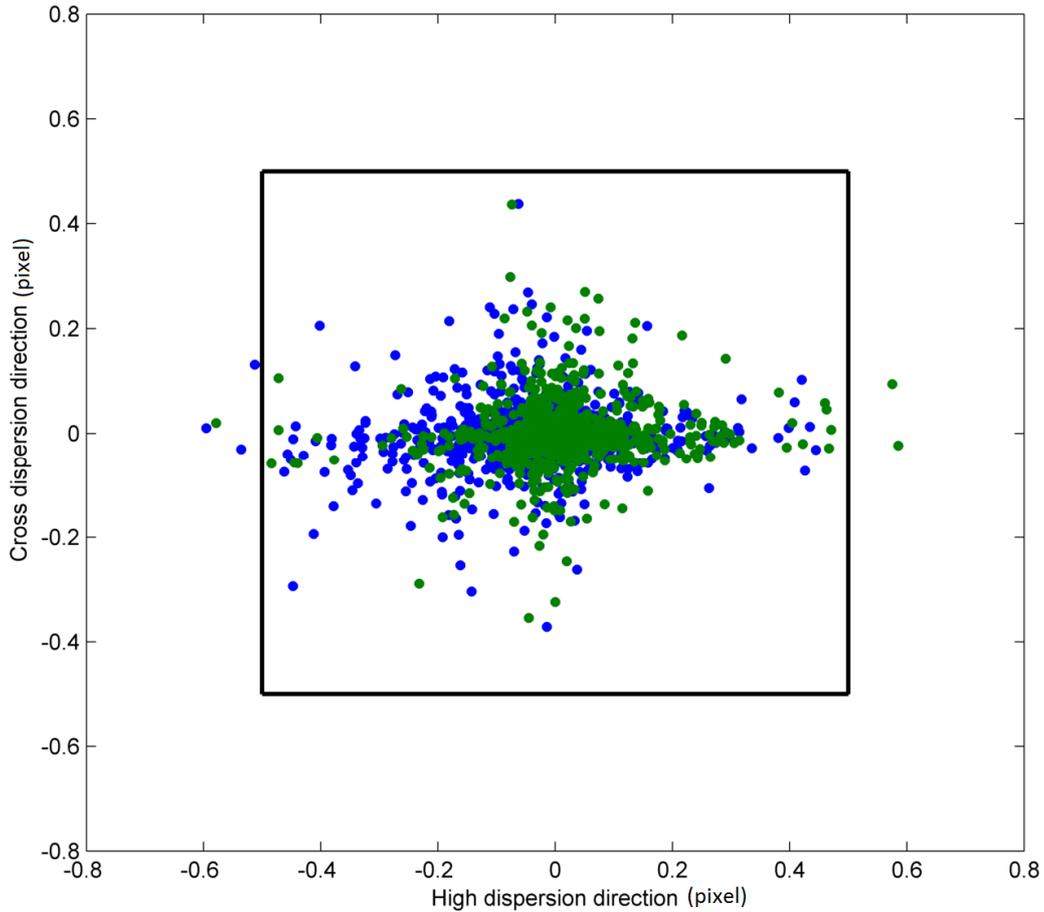

**Fig 4.4** Scatter of test features' residuals in position for model optimized using 233 features. Blue: Residuals before the polynomial fit to correct for the residual patterns seen in Fig 4.3, Green: Residuals after the 2D polynomial fit to correct the residual patterns seen in Fig 4.3

### 4.1 OUTLIERS

It is important to understand the reason for the outliers since a realistic model, good S/N lines and very small uncertainties in the wavelengths should not cause such outliers. More scatter can be observed in the high dispersion direction than the cross dispersion direction. The black lines shown in Fig 4.5 (High dispersion coordinate residuals for different test features) depict the 0.05 pixel and 0.1 pixel bounds. From visual observation, most of the lines are concentrated between the 0.05pixel bounds. About 80% of the residuals are between -0.1 pixel and 0.1 pixel, and 63% of lines residuals are between -0.05 and 0.05 pixel. An improper polynomial fit to the model is expected to show a trend in the residuals, hence the large residual and outliers could be due to issues in the centroid measurement of the Th features in the image or uncertainty in the wavelength of the feature considered for the model. The uncertainties in the Ritz wavelengths calculated by Redman et al.(2014) [4] cannot account for the outliers of this magnitude. About 30% of the outliers (marked blue in the Figure 4.5) are very weak lines with high errors in the centroid positions. The right end of the scatter plot corresponds to the blue end of the spectrum whose efficiency is generally poor. This could be the reason for significant scatter in this end. About 50% of the outliers have identified neighbours (Th or Ar) within less than one FWHM of the feature (Marked green in the Fig 4.5). When the blended features are identified as a single line and , the centroid is determined using a single Gaussian, then the measured centroid will be shifted towards the blended line. This shift is dependent not only on the separation between the lines but also the relative intensities of the lines. The rest of outliers were analysed individually. It is observed that while most of these lines have blends which are lines not identified in the atlases used, some lines have intense gradient back ground which can cause centroid shift. The FWHM values of the Gaussian fits and visual inspection were used to determine the possibility of a blend. The FWHM of the feature was compared to neighbouring features. Considerable discrepancies are concluded as presence of blends in the features. Mean and standard deviation of the residuals of the test features without these outliers improved to 0.000957 pixels and 0.04097 pixels respectively in the cross dispersion and high dispersion directions.

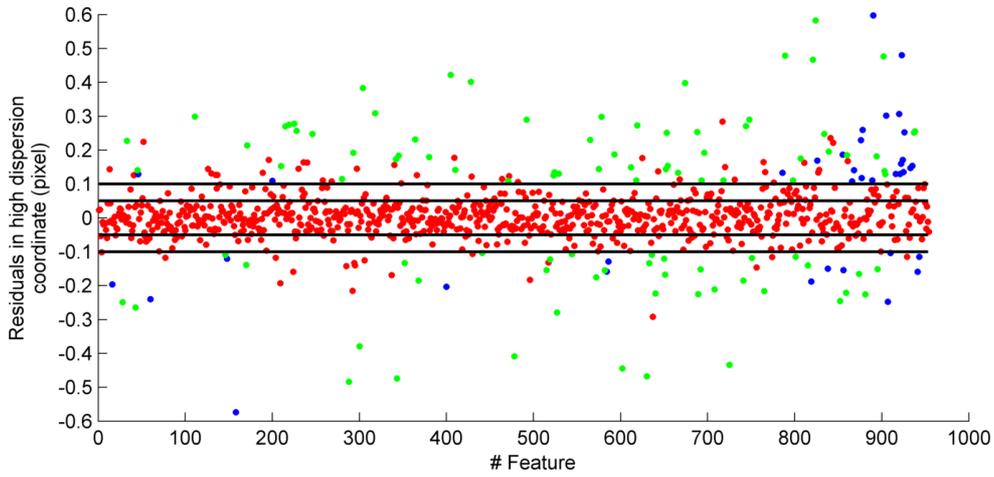

**Fig 4.5** Difference in high dispersion coordinates of measured and optimized model predicted positions of the test features for different features used. The features in the left are in the red region and the blue region is to the right. The two black solid lines depicts the 0.1 and 0.05 limits. Red dots: features in general, Green dots are features with blend/blends and Blue dots are weak intensity features with high errors in centroid measurements.

### 4.2 DISPERSION SOLUTIONS

As the next step, the dispersion estimate by the model was compared with conventional reduction and dispersion solution estimate using IRAF. The echelle reduction package in IRAF fits a 2D function to the ThAr dispersion relation. The two variables being the order and the high dispersion coordinate.

$$\lambda = f(x,o)/o$$ where o is the order number and x is the position coordinate.

$$f(x,o) = \sum_{n=0}^{opow} \sum_{m=0}^{xpow} C_{mn} P_{xm} P_{on}$$

Where opow and xpow are the maximum powers of x and o, and $P_{xm}$, $P_{on}$ can be Chebyshev or Legendre polynomials with x and o variables respectively. The order of x and o are chosen such that the optimum root mean square is achieved and yet the polynomial doesn't over fit the data. As it is not possible to do an inverse operation with the model, a grid of positions of various wavelength features in the spectrum are calculated from the optimized model. The so built grid is used for interpolation of unknown features. Fig 4.6 shows the scatter of the residuals for the two solutions for the same set of features. Open circles indicate the residuals of the dispersion solution from polynomial fitting of IRAF, while the squares indicate the residuals from the model as described above. The root mean square error has reduced by 10 times. The intensity in the spectrum falls considerably in the blue end which is visible as spread in scatter.

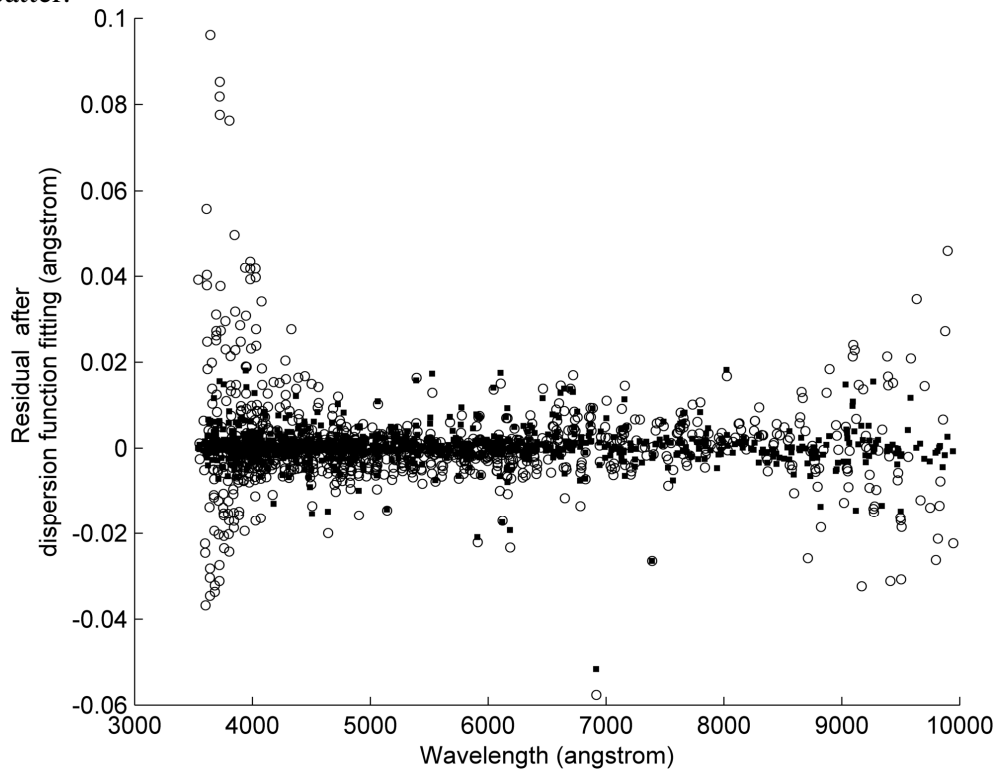

**Fig 4.6** Residuals around the dispersion solutions in high dispersion axis. Open circles are the residuals around IRAF based polynomial, solid square points are residuals in the model based technique.

### 4.3 NOISE PRECISION OF MODEL PREDICTIONS

Different lines in the spectrum have different intensities and accordingly different uncertainties associated with the positions measured. To understand the effect of these uncertainties on model optimization and predictions, bootstrap Monte Carlo simulations were performed. The coordinates of the lines used for model optimization were subjected to random variations with Poisson statistics of photon noise (as lines selected have negligible readout and bias noises compared to photon noise). These positions were used to optimize the model. The standard deviation of the predicted features was measured. Fig 4.7 shows the error in the measured positions and the error in predicted positions from the model. It is also observed that the error in model predictions is dependent on the intensity of lines used for optimization and the bounds used on the open parameters during optimization (given the possibility of multiple solutions from the optimization). A gradual change of uncertainties across the orders and a variation within an order itself can be observed (Fig 4.7 bottom panel). These are the direct results of the instrument behaviour for changes in various parameters.

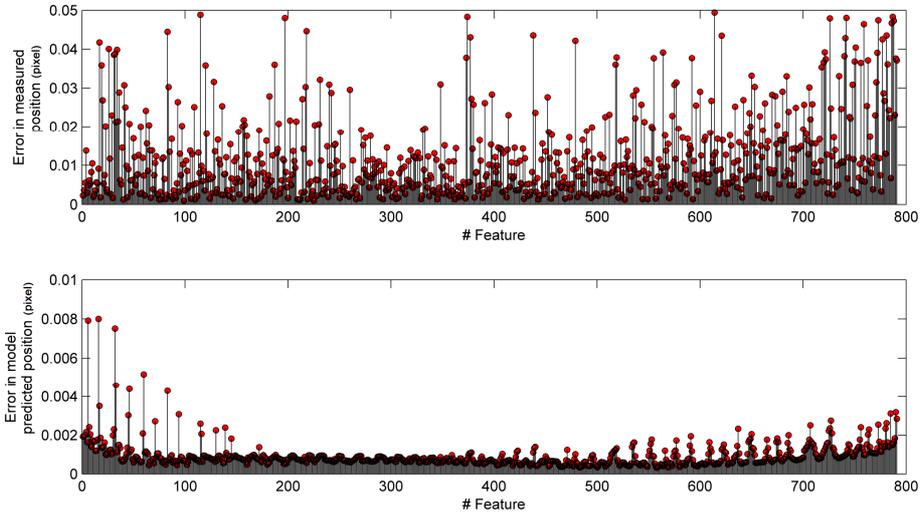

**Fig 4.7** Top panel: Error in measured positions of selected features in pixel. Bottom panel: Uncertainties in the model predictions of the same features from the photon noise in the features selected for optimization. The patterns seen in the bottom panel are the result of the instrumental behaviour for changing parameters

### 4.4 PARAMETERS

This subsection deals with the parameters that are optimized to match the model with instrument behaviour. As discussed in Section 2, degeneracy states among the parameters are a problem. While it is hard to measure the parameters to the required accuracy, in the following we describe how the bounds for parameters are fixed. There are two steps in using the model: Initial set of open parameters and optimization bounds on the open parameters 1. during system alignment, 2. during the scientific operations. During alignment the parameters will vary to the maximum, while during regular operations they vary to a very small extent around the parameters achieved from the alignment. The alignment procedures will give an idea of how much a particular parameter can vary. For example, the opto-mechanics' manufacturing tolerances can be accurate to about 100micron, input optics form the image of the fibres at the slit plate whose width and height are known, and the optics are aligned such that the images fall within the slit with proper clearance, different mechanical jigs were used to place the spectrograph subassemblies on the optical table to a high level of accuracy. It is also made sure that the open parameters never reach the bounds during optimization.

In order to understand the degenerate sets, optimization of the parameter set was carried out multiple times on same feature set. The so obtained optimized parameters were plotted and observed. Fig 4.8 shows plots for a few parameters. The parameters were in general varying around a mean value and the scatter around the mean value was mostly dependent on the bound levels chosen for the parameters that constitute the degenerate sets. Some parameters did not exhibit this behaviour as their bounds did not include the physical parameter value in the instrument. Such parameters always reached the bound value after optimization. Bounds were readjusted for such parameters and the model re-optimized. For the later stages of optimization, the mean value of these optimized parameters is made the initial parameter set and the bounds are set according to the deviations. The values and bounds proved sufficient for photon noise precision tests as well.

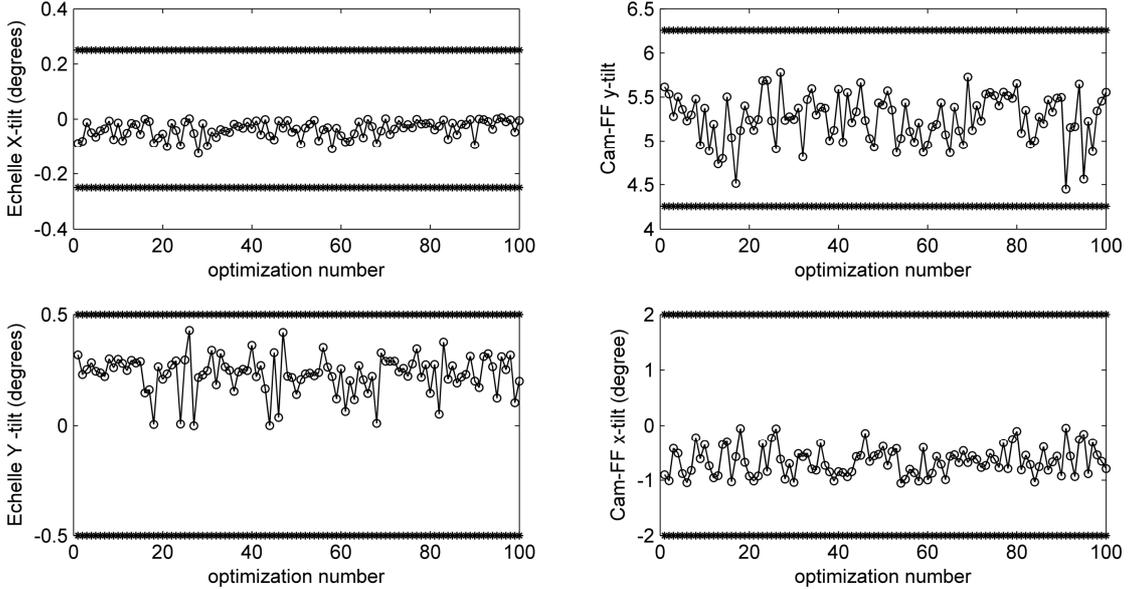

**Fig 4.8** Plots show the variation in parameters (X & Y-tilts of Echelle grating, X & Y tilts of the Camera1 and Field flattener interface) over a number of optimisation runs on the same set of features. The bold lines in each plot show the bounds used for respective parameters in the optimisation. The degeneracy in the parameter set causes the variations over different runs. It can be observed that the parameters vary about mean values which will form the initial parameter set during operations after the alignment

### 4.5 SPECTRUM SHIFT PREDICTION

With environment fluctuations and any vibrations at the instrument, the spectrum shifts on the CCD plane. For stellar radial velocity measurements, it is important to isolate the instrumental shifts from the stellar Doppler shifts. In order to determine the instrumental shifts, a ThAr spectrum is recorded along with the star spectrum. The shifts in the ThAr will provide the information about the instrumental shifts.

In this subsection we will understand the ability of the model to track these small shifts in the spectrum. ThAr exposures were taken over night to determine the instrument stability. The environment is stabilised over this time period with no external disturbances, except for small natural temperature and pressure changes, which was recorded. The conventional correction relies on using the whole ThAr spectrum to determine the shift, which is a single number over the entire spectrum or over an order. Cross correlation between a recorded spectrum and reference spectrum gives the shift. Environmental effects may not necessarily cause a constant radial shift throughout the spectrum as they tend to have a nonlinear effect on various parameters of the instrument.

The model was optimized for different spectra taken at different times over night and the shift through model predictions was calculated. In Section 4.3 the noise precision of the model predictions was shown. In principle any shift greater than these precision values can be detected. Fig 4.9 shows the shifts observed across the spectrum for two different observations with respect to a reference image.

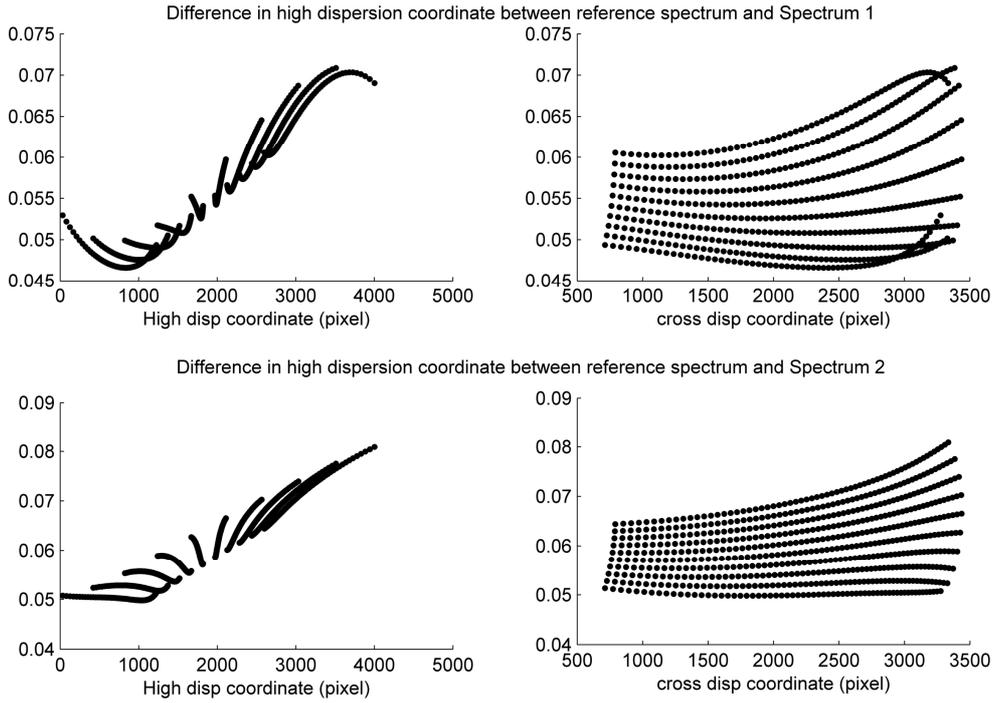

**Fig 4.9** Spectrum shift predictions by model. Shift in the high dispersion direction for two different spectra with respect to a reference spectrum as predicted by the optimized model.

It is observed that the pattern in the shifts vary from image to image. In order to confirm that the effect is not from the optimization procedure, multiple images were taken and a few features from them were chosen and the centroid differences were observed. It is observed that the shift indeed follow the pattern predicted by model as shown in Fig 4.10 as an example. The square marked error bars in the two plots (for two different orders) show the difference in positions of equally spaced features in an order predicted by model, while the circled error bar shows measurements from spectra obtained from the instrument.

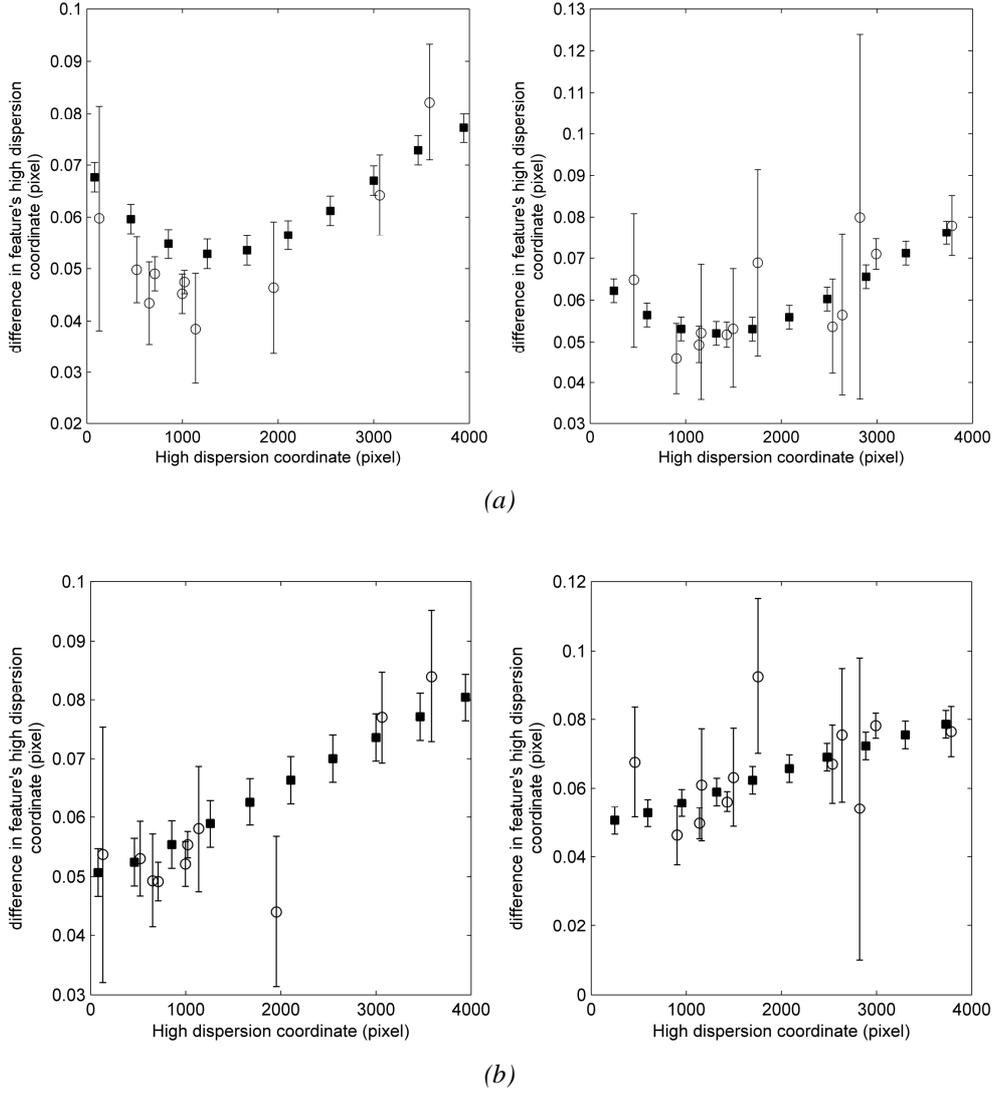

**Fig 4.10** Shift in spectrum in two different orders (along columns) with respect to a reference spectrum for two different images (a) and (b). Solid Squares are the model predicted shifts and the open circle are the measurements from the instrument exposures. This shows different patterns in the spectrum shifts in two different images taken at different points of time with about same mean shift and the performance of the models predictions for the same.

### 4.6 DOUBLE FIBER

In the double fiber mode of operation, while one fiber is fed by star light the other is fed with the ThAr light. The purpose is to record the instrumental drift in the ThAr spectrum and use it for correction in the science object spectrum. Two important issues in this operation are the difference in the high dispersion scale and the difference in shift between the two fibres. The slit length undergoes a curvature in the Echelle dispersion causing curved slit images on the CCD. Two fibres can be thought of as a long slit, which causes an angle between the two fibres on the CCD dependent on the wavelength. To compensate for this the input slit optics were rotated by a mean angle in the opposite direction. In order to determine the distance and angle between the two fiber images at the slit, exposures were taken feeding both the fibers with ThAr light. The parameter set optimized for one fiber is used and the decentre in x and y of the second fiber position are left open and re-optimized using the second fiber's features in the spectrum. The difference in high dispersion coordinates of the two fibre images at different parts of the spectrum were compared between the model predictions and features in the CCD image. The angle determined by the model matched the physical angle given to the input optics. Also the difference in centroids of the same features in the two fibres was compared between the model predictions and those measured from CCD images. Fig 4.11 shows an example of this difference for two different orders. It was observed that model is able to track the differences well over the entire spectrum.

In conventional data reduction and analysis, the shift in spectrum in the two fibres' channels is assumed to be the same though there is a slight difference in the paths traced by the light beam between the two.

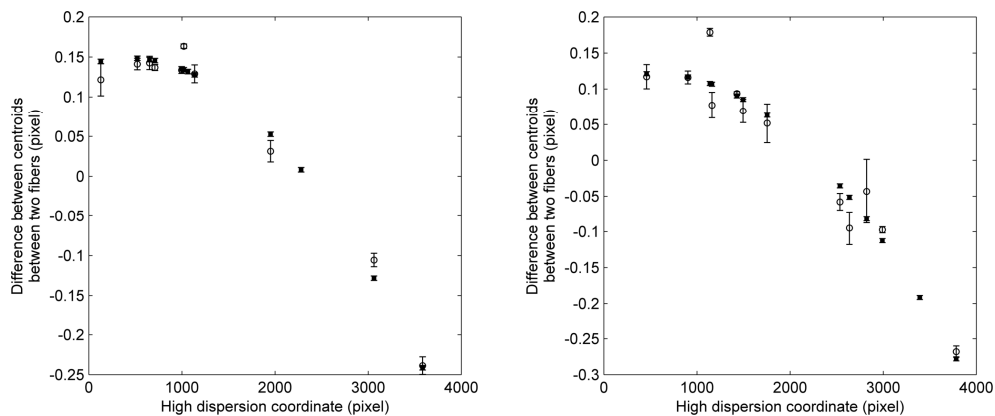

**Fig 4.11** Difference in high dispersion coordinates of centroids of the two fiber images in two orders in a spectrum (left and right subplots). Solid squares represent the model predictions and the open circles are the error bars for the measurements from the CCD exposures. Model optimized using one fiber spectrum is able to track the second fiber spectrum.

We tested if the model can predict the difference in shift between two channels. In different orders, the difference in shift of spectrum between the two fibres was measured and compared with model predictions. Fig 4.12 shows the same as an example for two different orders. The weighted mean of the shift difference between the two fibres in the ThAr calibration frames matched the mean shift value from the model. The model predictions were well within the errors of the measured shift differences and it is also observed that these differences in shifts between two fibres is varying over the spectrum.

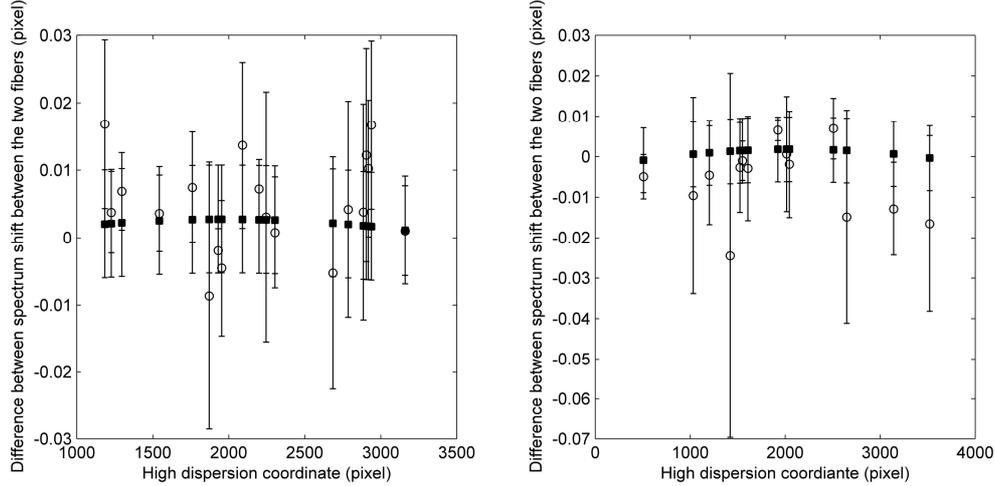

**Fig 4.12** Error bars showing the difference in the spectrum shifts between the two fibres for two orders (left and right subplots). Solid squares represent model predictions and the open circles are the measurements from the CCD exposures.

## 5 CONCLUSION

Building an accurate model is important to understand the final behaviour of the instrument without any ambiguities. For this, the best available values of built components were incorporated into the model along with accurate glass catalogues, environmental parameters, and wavelength atlases. The current optimization did not consider any weights for the parameters. Since we are dealing with physical parameters of a physical instrument, the effect of various parameters on the output can be studied quite well and accordingly they can be weighed for optimization.

The model was the tested using the ThAr calibration frames obtained during the HESP pre-shipment testing at the Kiwistar Optics lab. The strong Th features used for matching were selected such that there were 1-4 features in each order. Except for few outliers the mean and standard deviation of the residuals of test features are 0.000957 pixel and 0.04pixel respectively along the Echelle direction and even better in the cross dispersion direction. The dispersion solution of the model was compared with the 2D polynomial dispersion function from IRAF. The residual of the model fit was ten times better than the empirical fit using IRAF especially in the regions with very few features. The noisy data and varying efficiency of the instrument from red to blue region has lesser effect on the model based dispersion solution than the conventional empirical fit. The model based calibration technique works well in regimes where there are fewer calibration lines and an empirical calibration is less accurate. This is especially useful for the radial velocity studies of M-dwarfs, that has lines mainly in the red region and usable ThAr lines in this region is limited [5]. M-dwarfs are interesting for the better detectability of earth like planets with current instruments.

Accurate wavelength calibration using model based calibration in blue and red regions of visible wavelength will help in accurate abundances. The line features are broadened due to isotopic shifts and hyperfine structures. Several high resolution spectrographs use image slicers for high throughput at high resolution [6][7][8]. These spectrographs have typically 2-pixel sampling. During a 1D spectral extraction the pixels along the slit length are added to achieve high S/N. The procedure assumes the added pixels have same wavelength. However tilt along the spatial direction will cause cross talk and the isotopic abundances derived from such extraction will be inaccurate. A model can give a complete 2D wavelength map of the detector that will help in accurate extraction, hence accurate abundances. In the red region of the optical spectrum there are several telluric absorption lines and sky emission lines. Accurate sky subtraction and telluric line removal is needed to derive accurate abundances of some of the important stellar atomic lines, (e.g. HeI, CI, NI, OI, KI and SI) [9][10][11]. Accurate removal of sky lines and telluric lines needs good wavelength calibration.

It is also observed that for the same mean instrumental shift observed in the spectrum, the pattern of shift across the spectrum may vary. The model was able to predict these patterns. This sort of information can be useful in applying position/wavelength dependent instrumental drift corrections and also extending the drift corrections to areas where the calibration lines are sparse and weak.

For double fiber observations, the model was able to track the difference in the centroids between the two fiber images across the spectrum. The model was optimized using only one fiber spectrum, yet it was able to track the centroid differences between the two fibers which is varying across the spectrum. This is an indication that the optimized model can map the slit position onto the CCD plane. Using this, an efficient 1D extraction of the spectrum can be performed for varying slit image tilts on the detector plane.

We also attempted to test if the model is able to predict the difference in the instrumental shifts between the two fibres. The mean values of these differences matched with the mean values calculated from the model and also the difference in shifts between the fibres varied over the spectrum.

The double fiber, simultaneous referencing method relies on the assumption that the reference calibration tracks any changes to the object spectrum. The difference between the object and the reference fiber is assumed to be a mean instrument drift that can be added to the dispersion relation derived at the beginning of the night. In a real case, the instrument drift is a function of position and wavelength. In future instruments ,when we are not limited by photon noise, the positional dependence of instrument drift will play an important role [12]. This helps in using the entire wavelength coverage of the spectrograph, for precise radial velocities. This also helps in accurate sky subtraction when the second fiber is used for collecting the sky light, especially beyond 600nm where there are sky emission lines due to $O_2$, $H_2O$ [13]. Sky fibers are used to remove the contribution from these lines. Inaccurate determination of instrumental drift between the two fibers, especially the wavelength dependency will result in inaccurate line strengths and chemical abundances. Sulphur is one such important element, which occurs in the crowded regions of telluric and skylines. Sulphur lines are very important, since this is one of the heavy alpha element diagnostics that is not affected by dust depletion [9]. Hence, Sulphur is a better indicator to study the planet frequency and metallicity correlation of the host stars [14], since sulphur probes the metallicity of primordial cloud. Similarly, the chemical abundances comparison of old metal poor stars and the high redshift inter-galactic medium studies are more meaningful for volatile elements like C,N,O and S , that are less affected by dust in the high red shift intergalactic medium [15]. These lines occur in the red region from 700-1000nm, which is severely affected by sky and telluric lines. While these are the expected application areas for the model based calibrations from the performance results so far, the model's application to the science cases is underway after the installation of HESP at the telescope, which will be reported in future publications.